# A FRAMEWORK FOR THE ADOPTION OF BIOMETRIC ATM AUTHENTICATION IN THE KENYAN BANKS

**Benson M. Onywoki**

bensonmakori@ymail.com

**Dr. Elisha T. Opiyo**

opiyo@uonbi.ac.ke

*School of Computing and Informatics, University of Nairobi*

## Abstract

The use of ATMs has become fundamental in the banking industry owing to the values transacted using these systems and their 24/7 usage.

Although several researchers have studied the role of biometrics in security applications for financial institutions, no systematic empirical research has been applied to studying the role of organizational characteristics and contextual factors in the Kenyan financial sector.

This study sought to develop a framework for the adoption of biometric ATMs in the Kenyan banking sector, apply the developed framework to study factors influencing adoption of biometric ATM authentication and validate the developed conceptual framework.

A survey was used to collect quantitative data from the ATM users which was then analysed using factor analysis and multiple regression analysis. The study established that performance expectancy, effort expectancy, social influence and user privacy were key determinants for biometric ATMs acceptance, adoption and usage. The study further demonstrated that age, gender and experience were moderating factors on effort expectancy with experience further moderating performance expectancy, effort expectancy, social influence and user privacy.

# Introduction

Banking in Kenya has come a long way from the time of ledger cards and other manual filling systems. Most banks today have electronic systems to handle their daily voluminous tasks of information retrieval, storage and processing. Irrespective of whether they are automated or not, banks in Kenya by their nature are continually involved in all forms of information management on a continuous basis.

Among the results of increased number of bank account holders were the long queues at the banking halls. Banks had no option but to adopt the Automatic Teller Machines, ATMs, that were to handle transactions which included cash withdrawals, mini statement requests, balance inquiry and later cash and cheque deposits.

More than 45 years after their invention, ATMs are being used to perform a number of functions, ranging from traditional cash dispensing, cash deposits, account transfers, mini statements and even payment of bills. The adoption of the ATMs by the banks was a great milestone. It reduced queues in the banking halls and customers could access their accounts any time and day of the week.

The use of ATMs in the Kenyan banking sector has not failed to encounter challenges. Security of information and transactions ranks high among the challenges that countries and organizations continue to grapple with. As efforts to address these challenges continue to gather momentum, focus is quickly shifting to information technology to provide a solution.

Accurate and automatic identification and authentication of users is a fundamental problem in network environments. Personal Identification Numbers or Passwords and key devices like Smart cards are not just enough in some cases. What is needed is something that could verify that you are physically the person you claim to be (Koteswari et al, 2011). The use of biometric authentication has been identified as a solution for this problem.

Substantial interest exists in biometric authentication technology and its application as a means of enhancing existing identity management systems, but there are also valid concerns when protecting the stored data (Wayman, 2008). This poses a principal adoption decision for an organization because this type of authentication requires strict privacy measures, adequate database management, and a long-term financial commitment due to the sensitive nature of the data being stored (Laux, 2011).

Because biometric technologies encompass a group of technologies designed to identify and validate the identity of individuals using one or more of their intrinsic and unique physical or behavioral traits (Bolle et al, 2004), it represents a robust technology for deterring fraudulent access to online, as well as on-site, financial services.

Breckenridge (2005) examined the deployment of biometric technology in South Africa. He suggests that before an organization embarks on an implementation plan of biometric system deployment, it would be wise to examine the lessons that can be learned from South Africa's troubled adoption of a biometric-based authentication system.

## Literature Review

Below are some definitions of the key terms used in this research

*Biometrics* - This is a science and technique for recognizing the human characters, both physiological and behavioral. (Gaur et al, 2012)

*Authentication* - Corcoran et al., (2000) describes authentication as the process of associating an individual with his/her' unique identity, which is the manner he/she establishes the validity of his/her claimed identity.

*Model* – this is a conceptual structure intended to serve as a guide for developing something that expands the structure into something useful.

*Adoption* – means to choose as a standard or required in a course.

*Effort expectancy:* The degree of ease associated with the use of the system.

*Attitude:* Individual's positive or negative feeling about performing the target behavior (e.g., using a system).

*Behavioral intention:* The degree to which a person has formulated conscious plans to perform or not perform some specified future behavior.

*Performance expectancy:* The degree to which an individual believes that using the system will help him or her to attain gains in job performance.

*Perceived usefulness:* The degree to which an individual views the system in relation to helping him or her attain gains in job performance.

*Subjective norm:* Person's perception that most people who are important to him think he should or should not perform the behavior in question. Subjective norm is also referred to as social influence.

**Technology Adoption Models**

Information technology adoption by users is considered to be attributed to various drivers or factors that are interrelated. Various models have been put forth to facilitate the understanding of Information Technology adoption. Below, we discuss the theoretical models and illustrate the key constructs of IT adoption that each model addresses and relate them to adoption of biometric authentication.

*The Theory of Reasoned Action (TRA)*

Applied to adoption of biometric authentication in the banks, TRA seems to maintain that, individuals would accept biometric authentication if they could see that there would be positive benefits or outcomes associated with using that technology (Fishbein et al, 1975). The TRA model is a widely studied model from social psychology concerned with the determinants of consciously intended behaviors. However the model does not pay tribute to other adoption drivers such as facilitating conditions offered by new technology, performance expectancy and effort expectancy as well as cost. The model mostly attributes importance on social influence on intended behavior to adopt a system.

*Theory of Planned Behavior (TPB) Model*

The Theory of Planned behavior asserts that individual behavior is driven by behavioral intentions. TPB improves the predictive power of the theory of reasoned action by including perceived behavioral control. This could imply that an individual does not have a complete control of their behavior under certain conditions. Perceived Behavioral Control refers to the perceived ease or difficulty of performing or exhibiting behavior (Ajzen, 1980).

*Social Cognitive Theory (SCT) Model*

In this model, behavior, Personal, and Environmental influences operate interactively as determinants of each other.

*Unified Theory of Acceptance and Use of Technology*

UTAUT considers four constructs hypothesized to have a significant role as direct determinants of user acceptance and usage behavior. The four constructs include; performance expectancy, effort expectancy, social influence and facilitating conditions.

The constructs are moderated by age, gender, experience and voluntariness of use to determine user acceptance and usage behavior (Venkatesh et al., 2003)

**Summary of the Gaps in the Reviewed Adoption models**

There are some shortfalls associated with the use of the discussed models in determining the adoption drivers of biometric authentication. Table1 summarizes these weaknesses.

Table 1: Summary of the Gaps in the Reviewed Adoption models

| Model | Gaps |
|---|---|
| The Theory of Reasoned Action (TRA) | Does not pay tribute to other adoption drivers such as facilitating conditions offered by new technology, performance expectancy and effort expectancy as well as cost. |
| The Theory of Planned Behavior (TPB) | TPB focuses more on users' behavior towards adopting a system and does not consider design and implementation strategies of the system to be adopted. |
| Social Cognitive Theory | Because social cognitive theory is so broad, it has been criticized for lacking any one unifying principle or structure. |
| Unified Theory of Acceptance and Use of Technology | UTAUT has not addressed the user privacy concern which is a key factor in the adoption of the biometric authentication systems. |

**The conceptual framework**

This study finds the Unified Theory of Acceptance and Use of Technology, UTAUT, as a possible model to extend in order to arrive at the framework for adoption of biometric authentication.

UTAUT has consolidated constructs from various models to come up with a single model which addresses most of the adoption factors.

This research therefore identifies five factors that affect the adoption of the biometric authentication in the Kenyan banks namely:

1. Performance expectancy
2. Effort expectancy
3. Social influence
4. Facilitating conditions
5. User privacy

*Performance expectancy*

This refers to the degree to which the stakeholders believe that using biometric authentication at the ATMs will help them improve transaction security.

*Effort expectancy*

Effort expectancy refers to the degree of ease associated with the use of biometric authentication at the cash dispensing machines.

*Social influence*

This is the importance to which an individual attaches others' beliefs on the use the biometric authentication systems at the bank ATMs.

*Facilitating conditions*

This is the degree to which an individual believes that the Kenyan banks and technical infrastructure exists to support use of system or service. The compatibility of the new biometric systems with the current implementations is also considered here. This factor also looks at the cost of implementation, whether the bank directors and shareholders will be willing to fund the implementation of the biometric authentication ATMs.

*User privacy*

For the purposes of our study we will consider user privacy as a factor for the adoption of Biometric authentication in the Kenyan banks. The concerns on privacy of the biometric data during the enrollment stage and its storage are major issues that can affect the adoption of this technology. It is therefore important to know to which extent the customers and the regulators are concerned about this factor.

Biometric data contains information acquired from individuals, which can be used to identify them. This raises issues of privacy and data protection. If the biometric data is recorded in a central database, privacy concerns may be higher than for systems where an individual's data is stored only on a card retained by the individual. Note however, some biometric applications require a central database for their basic functionality e.g. to check for multiple enrolment attempts.

Customers may be concerned that their biometric data could be used for purposes other than it was originally acquired; for example, face image data might be used for surveillance purposes and fingerprint data checked against forensic databases. These concerns are at the heart of many objections to the use of biometrics. It is therefore necessary to understand privacy issues in regard to biometric data and biometric systems.

## Methodology

This research used a survey technique in the collection of data. The researcher used questionnaires as the instrument of data collection. The design of the research instrument consisted of two sections; part one covered the demographic characteristics of the respondents and part two covered the empirical measurements for the constructs in the proposed model.

The determinants in the proposed framework were: Performance expectancy, Effort expectancy, Social influence, Facilitating conditions and User Privacy.

## Findings and Discussions

Factor analysis was run on the sample. The extraction method used was principal component analysis (PCA) with varimax rotation method.

Four of the conceptual framework variables are supported by results of factor analysis. These variables include **Performance expectancy, Effort expectancy, Social influence and User privacy**.

Facilitating conditions was not found to be a factor in the adoption of biometric ATMs.

**Testing for Direct Effects**

Regression analysis was used to test the relationship between the four constructs and the adoption of biometric ATM authentication.

Regression analysis is a statistic technique used to investigate the relationships between a dependent variable and one or more independent variables. Multiple linear regression is used in this study investigate the relationship between the behavioral intention and the four independent variables.

Regression coefficients can be used to evaluate the strength of the relationship between the independent variable and the dependent variable. $R^2$ value provides a measure of the predictive ability of the model. The close the $R^2$ value to 1 the better the regression equation fit to the data.

The F test is used to test the significance of the regression model as a whole. F is a function of $R^2$, the number of independent variables and the number of cases. F is computed with k and (n-1) degrees of freedom, where k = number of terms in the equation not counting the constant (Garson, 2008). The decision rule for F-ratio is to reject the null hypothesis if F is greater than the critical value of an appropriate level of significance, and not to reject the null hypothesis when F value is smaller or equal to the critical value of an appropriate level of significance.

Multiple regression analysis was used in this study to test the research hypothesis.

The regression model can be presented as follows;

**BI=a+b1PE+b2EE+b3SI+b4UP +e**

Where: BI=behavioral intention
PE=performance expectancy
EE=effort expectancy
SI=Social Influence
US=User Privacy
a= the constant where regression intercepts the y axis
b= regression coefficients and e = random error

The four independent variables, Performance expectancy, effort expectancy, social influence and user privacy were regressed against behavioral intention and provided the results in the table below.

**Table 2: Testing of direct effects**

| Model | Unstandardized Coefficients | | Standardized Coefficients | t | Sig. |
|---|---|---|---|---|---|
| | B | Std. Error | Beta | | |
| performance expectancy | 0.012 | 0.036 | 0.018 | 0.002 | 0.032 |
| Effort expectancy | 0.128 | 0.046 | 0.176 | 2.720 | 0.192 |
| Social influence | 0.007 | 0.038 | 0.012 | 0.115 | 0.373 |
| User Privacy | 0.013 | 0.020 | 0.028 | 0.422 | 0.171 |

All independent variables obtained positive beta weights hence have positive effect on the adoption of Biometric ATM authentication.

**Testing Moderating Effects**

With respect to interaction variables, the relationships are measured by Beta values, which represent the strength of the relationship. The Beta for the interaction of the moderator with the variable provides information regarding the interaction effect.

The Beta values should not be less than 0.1 and if they go beyond 1, there is a sign of Multicollinearity.

**Table 3: The moderating effect of gender, age and experience**

| GENDER | ß | AGE | ß | EXPERIENCE | ß |
|---|---|---|---|---|---|
| PE*GENDER | -0.037 | PE*AGE | -0.005 | PE*EXPERIENCE | 0.107 |
| EE*GENDER | 0.100 | EE*AGE | 0.133 | EE*EXPERIENCE | 0.180 |
| SI*GENDER | -0.028 | SI*AGE | -0.004 | SI*EXPERIENCE | 0.156 |
| UP*GENDER | -0.018 | UP*AGE | 0.005 | UP*EXPERIENCE | 0.144 |

According to the results above, gender has no effect on performance expectancy (beta=-0.037), Social influence (beta=0.028) and user privacy (beta=0.018). It however had a slight effect on effort expectancy (beta=0.100).

Age has beta values of -0.005 on performance expectancy, -0.004 on social influence, and 0.005 on user privacy. Age only moderates effort expectancy with a beta value of 0.

Experience has beta values of 0.107 on performance expectancy, 0.180 on effort expectancy, 0.156 on social influence and 0.144 on user privacy. Experience therefore moderates all these factors since it has beta values more than 0.100. Figure 1 below presents a validated framework for Biometric ATM Authentication

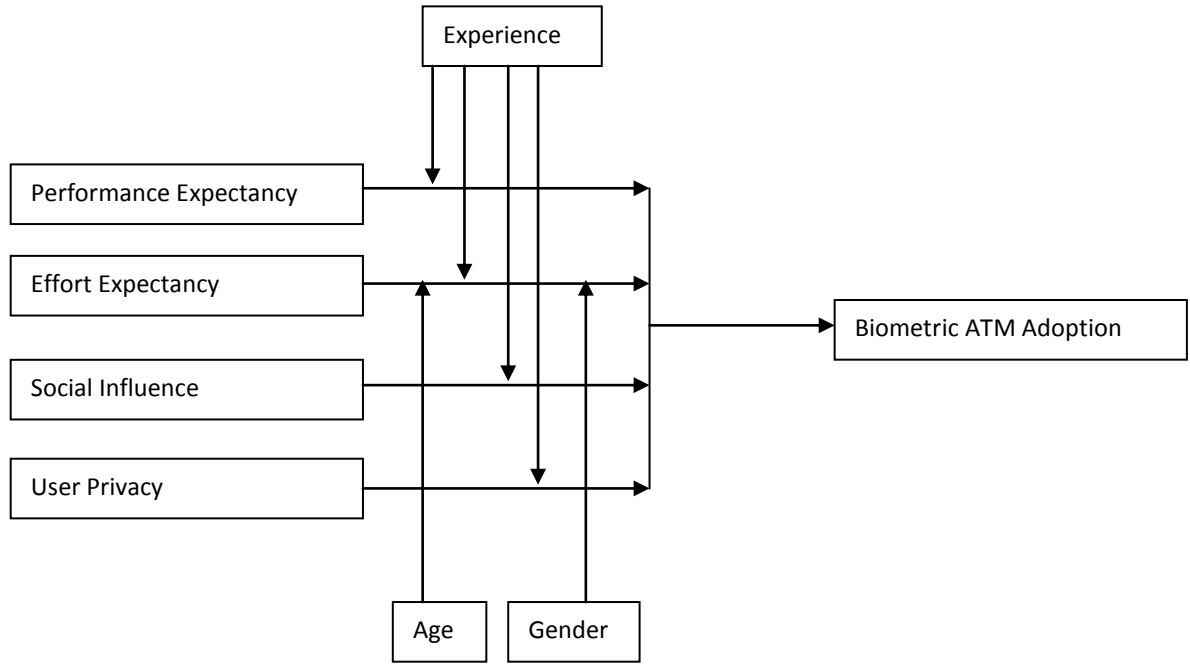

**Figure 1: Validated Framework for Biometric ATM Authentication**

# Conclusion

This research developed a framework which was validated by the findings. Based on this research framework, the study established that performance expectancy, effort expectancy, social influence and user privacy were key determinants for biometric ATMs acceptance, adoption and usage in the Kenyan Financial service providers. In an organizational setting, the management needs to increase performance expectancy, reduce effort expectancy, increase the 'hype' of using biometric ATMs and assure the customers that the biometric templates will only be used for the purpose that they were intended to. Facilitating conditions were not found to be having any effect on behavioral intention to use biometric ATMs.

The study further demonstrated that age, gender and experience were moderating factors on effort expectancy. Experience was a moderating factor on performance expectancy, effort expectancy, social influence and user privacy.

**Research Limitations**

Venkatesh et al (2003), points out that acceptance models examine technology from the time of their initial introduction to stages of greater experience. The responses in this research have been collected and examined to measure perceptions and expectations after the participants' acceptance or rejection decisions rather than during active biometric ATM adoption decision-making process.

**Contributions to Knowledge**

This research expands knowledge in the area of biometric applications' adoption and usage. Identifying the adoption drivers of such applications helps researchers and stakeholders to design training, marketing and infrastructure support to encourage biometric applications acceptance. The ATM networks for the banks are spread all over the country. For the current ATMs to be replaced much investments have to be made and if they are to recoup these investments, then the biometric ATMs must be accepted and be used by the customers. Only by understanding the barriers to user acceptance of this technology can the banks reduce those impediments.

This research identifies factors that are likely to affect the adoption of biometric ATM systems in the Kenyan banking sector. A clear understanding of these determinants will enable financial service providers to develop suitable marketing strategies, business models, processes, awareness programs and pilot projects

In conclusion, this study has contributed to knowledge with respect to theoretical extension and practical implementations. The validated framework can be further developed and refined to benefit other financial institutions that require user authentication.

**Recommendations for Future Research**

The application of biometric technology is wide. There are several types of biometrics that can be used at the ATM. The most common are finger print recognition, finger vein recognition, iris recognition, biometric hand geometry and facial recognition. This research focused on the general adoption of biometric ATMS and did not narrow down on the specific type of biometric authentication that can be used by the Kenyan banks. There is therefore need to future researchers to find out the specific type of biometric that will be appropriate to the users.

End-User Security Systems, *Journal of Organizational Computing and Electronic Commerce,* vol. pg 221–245.

Taylor S. and Todd P., 1995b. Understanding Information Technology Usage. A Test of competing models, *Information systems Research* 6(2): 144-176

Tomi D., Niina M., Anssi O., 2008, Trust enhanced technology acceptance model- consumer acceptance of mobile payment solutions
Venkatesh, V., Morris, M., Davis, G., Davis F. (2003) User acceptance of information technology: toward a unified view. *MIS Quarterly*, (27)3, pp. 425-478.

Wayman, J. L. 2008, Biometrics in identity management systems, *IEEE Security and Privacy*
*Magazine* 6(2) 30–37.
12